\documentclass[aps,prl,twocolumn,a4paper]{revtex4}
\usepackage{graphicx,psfrag}
\begin{document}

\title{Energy pumping in a quantum nanoelectromechanical system}
\author{T. Nord}
\email[]{nord@fy.chalmers.se}
\author{L.Y. Gorelik}
\affiliation{Department of Applied Physics, Chalmers University of
  Technology \\ and G\"oteborg University, SE-412 96 G\"oteborg,
  Sweden}
\date{\today}

\begin{abstract}
  Fully quantized mechanical motion of a single-level quantum dot
  coupled to two voltage biased electronic leads is studied. It is
  found that there are two different regimes depending on the applied
  voltage. If the bias voltage is below a certain threshold (which
  depends on the energy of the vibrational quanta) the mechanical
  subsystem is characterized by a low level of excitation. Above a
  threshold the energy accumulated in the mechanical degree of freedom
  dramatically increases. The distribution function for the energy
  level population and the current through the system in this regime
  is calculated.
\end{abstract}


\maketitle

During the past few years experimental methods of physics has seen
an advancing capability to manufacture smaller and smaller
structures and devices. This has lead to many new interesting
investigations of nanoscale physics. Examples include, for
instance, observation of the Kondo effect in single-atom
junctions\cite{a:02_park}, manufacturing of single-molecular
transistors\cite{a:02_liang}, and so on. There has also been a
great interest in the promising field of molecular
electronics\cite{a:00_joachim}. One of the main features of the
conducting nanoscale composite systems is its susceptibility to
significant mechanical deformations. This results from the fact
that on the nanoscale level the mechanical forces controlling the
structure of the system are of the same order of magnitude as the
capacitive electrostatic forces governed by charge distributions.
This circumstance is of the utmost importance in the so called
electromechanical single-electron transistor (EM-SET), which has
been in focus of recent research. The EM-SET is basically a double
junction system where the additional (mechanical) degree of
freedom, describing the relative position of the central island,
significantly influences the electronic transport. Experimental
work in relation to EM-SET structures range from the
macroscopic\cite{a:99_tuominen} to the micrometer
scale\cite{a:02_scheible,a:01_erbe,a:98_erbe} and down to the
nanometer scale\cite{a:00_park}. Various aspects of electronic
transport in such systems have been theoretically investigated in
a series of
articles\cite{a:02_nishiguchi,a:02_armour,a:02_mozyrsky,a:02_nord,a:02_fedorets,a:01_nishiguchi,a:01_isacsson,a:01_boese,a:01_gorelik,a:99_weiss}.

In \mbox{Ref. \onlinecite{a:98_gorelik}} and \mbox{Ref.
  \onlinecite{a:02_fedorets}} it was, among other things, shown that
coupling the mechanical degree of freedom of an EM-SET to the
nonequilibrium bath of electrons constituted by the biased leads, can
lead to dynamical self excitations of the mechanical subsystem and as
a result bring the EM-SET to the shuttle regime of charge transfer.
This phenomena is usually referred to as a shuttle instability. In
these papers the grain dynamics are treated classically and the key
issue is that the charge of the grain, $q(t)$, is correlated with its
velocity, $\dot{x}(t)$, in a way so that the time average,
$\overline{q(t)\dot{x}}\neq0$.

Decreasing the size of the central island in the EM-SET structure to
the nanoscale level results in the quantization of its mechanical
motion. Charge transfer in this regime was studied theoretically in
\mbox{Ref. \onlinecite{a:02_mccarthy}}. However, the strong additional
dissipation in the mechanical subsystem suggested in this paper keeps
the mechanical subsystem in the vicinity of its ground state and
prevents the developing of the mechanical instability. The aim of our
paper is to investigate the behavior of the EM-SET system in the
quantum regime when its interaction with the external thermodynamic
environment generating additional dissipation processes can be partly
ignored in such a manner that the mechanical instability becomes
possible. We will show that in this case at relatively low bias
voltages, intrinsic dissipation processes bring the mechanical
subsystem to the vicinity of the ground state.  But if the bias
voltage exceeds some threshold value, the energy of the mechanical
subsystem, initially located in the vicinity of the ground state,
starts to increase exponentially. We have found that intrinsic
processes alone saturate this energy growth at some level of
excitation. The distribution function for the energy level population
and the current through the system in this regime is calculated.

\begin{figure}[htb]
  \begin{flushleft}
    (a) \\
  \end{flushleft}
  \psfrag{ev}[]{\hspace*{0mm}\raisebox{0mm}{\large{$\mathbf{\frac{eV}{2}}$}}}
  \psfrag{hw}[]{\hspace*{0mm}\raisebox{0mm}{\large{$\mathbf{\hbar \omega_0}$}}}
  \includegraphics[scale=0.35]{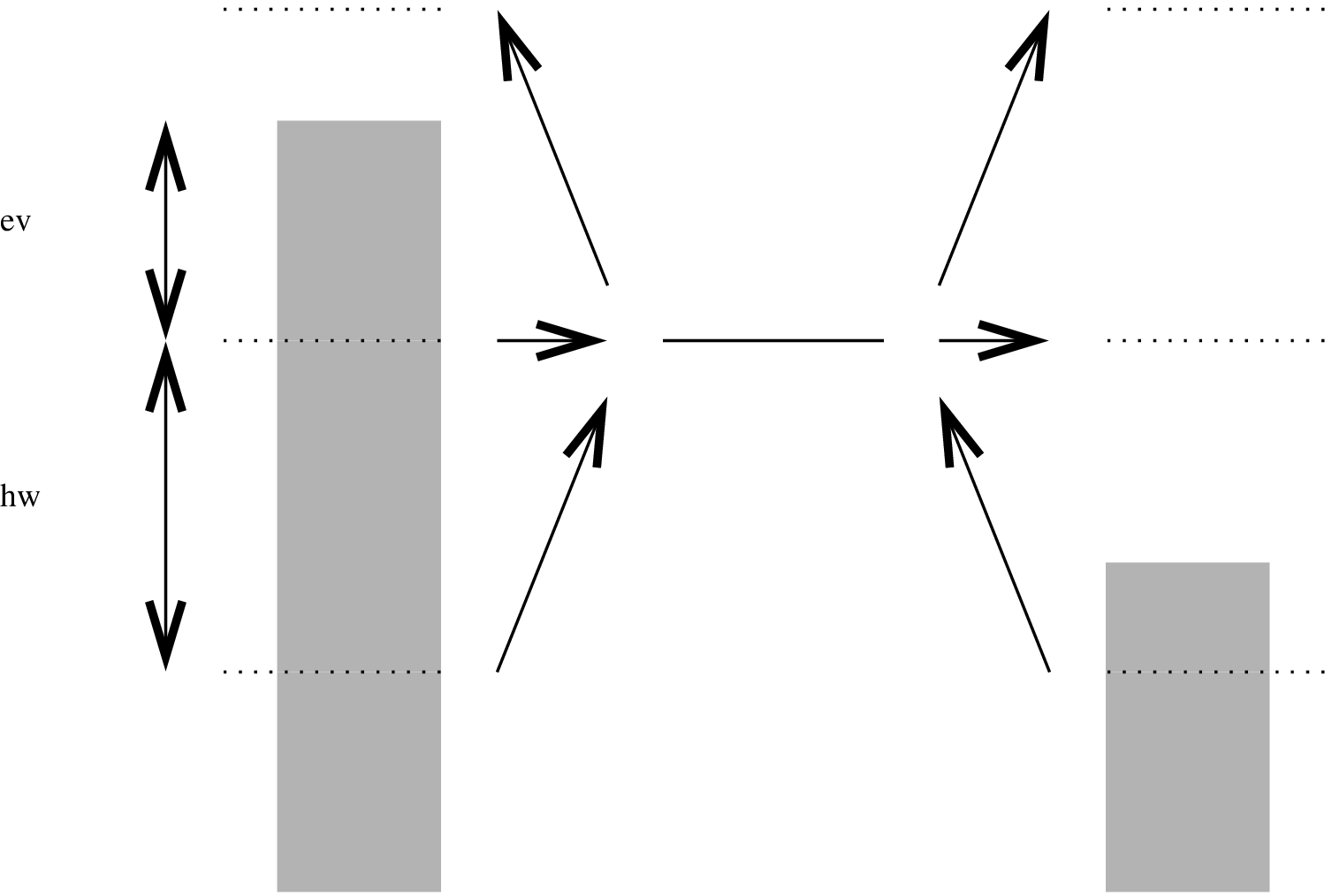}\\
  \begin{flushleft}
    (b) \\
  \end{flushleft}
  \includegraphics[scale=0.35]{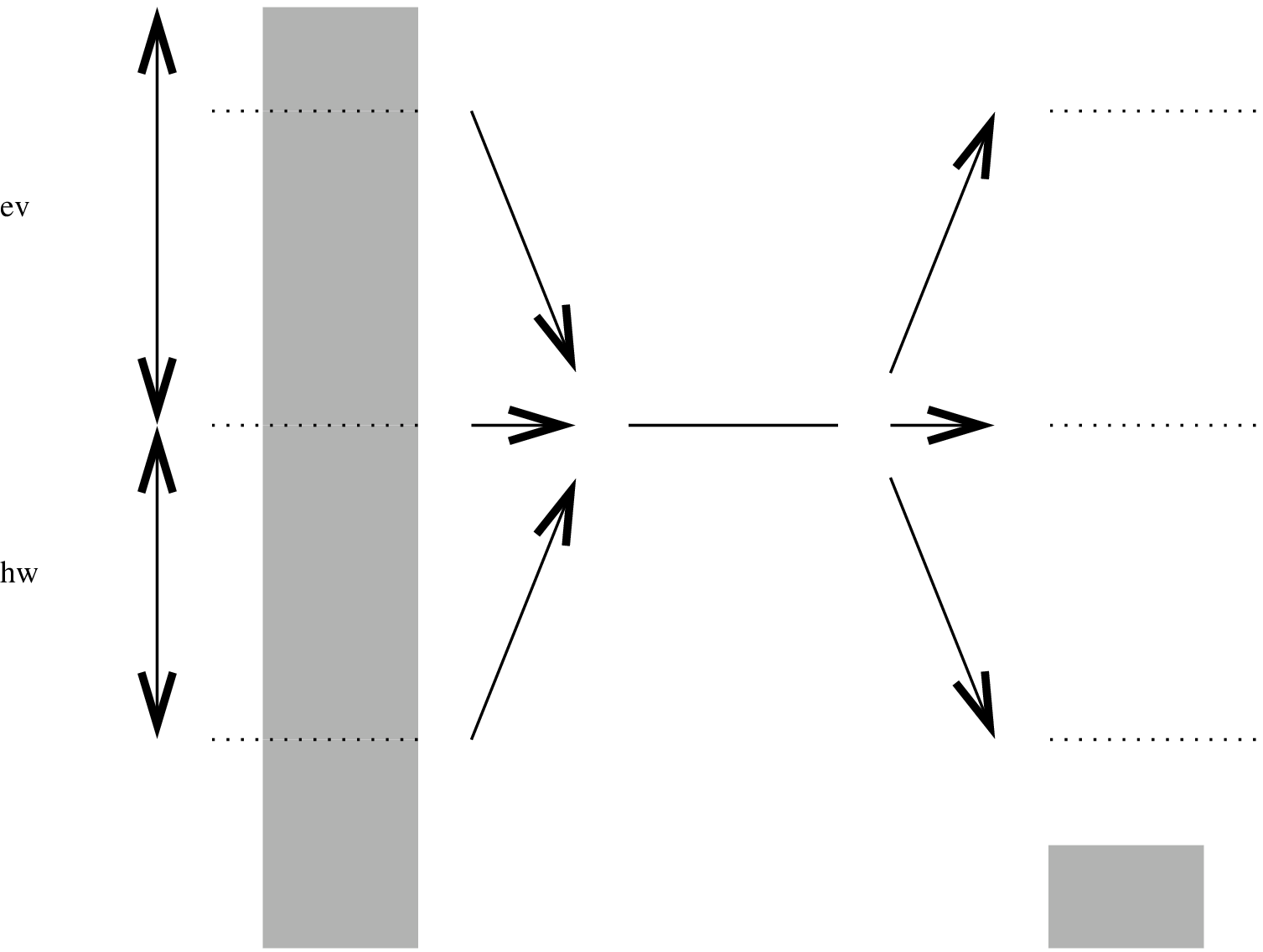}
  \vspace{5 mm}
  \caption{ \label{fig:case12}(a) Model system consisting of a one
    level quantum dot placed between two leads. The level of the dot
    equals the chemical potential of the leads and a bias voltage of
    $V$ is applied between the leads. The center of mass movement of
    the dot is in a harmonic oscillator potential with the vibrational
    quanta $\hbar \omega_0$. The applied bias voltage is such that
    $eV/2 < \hbar \omega_0$. (b) Same as (a) but with the applied bias
    voltage larger than $2\hbar \omega_0 / e$.}
\end{figure}

We will consider a model EM-SET structure consisting of a one
level quantum dot situated between two leads (see \mbox{Fig.
  \ref{fig:case12}}). To describe such a system we use the Hamiltonian
\begin{eqnarray}
  H &=& \sum_{k,\alpha} E_{k,\alpha} a_{k,\alpha}^{\dag}
  a_{k,\alpha} + (E_0 - D\frac{\hat{X}}{x_0})c^{\dag}c
  + \frac{1}{2m}\hat{P}^2 \nonumber \\
  & & + \frac{1}{2}m\omega_0^2\hat{X}^2
  + T_{k,\alpha}(\hat{X}) [ a_{k,{\alpha}}^{\dag}c
  + c^{\dag}a_{k,{\alpha}}].
  \label{eq:hamiltonian}
\end{eqnarray}
The first term describes electronic states with energies
$E_{k,\alpha}$ and where $a_{k,\alpha}^{\dag}$ ($a_{k,\alpha}$) are
creation (annihilation) operators for these noninteracting electrons
with momentum $k$ in the left ($\alpha=L$) or right ($\alpha=R$) lead.
The second term describes the interaction of the electronic level on
the dot with the electric field so that $c^{\dag}$ ($c$) is the
creation (annihilation) operator for the dot level electrons and $E_0$
is the dot energy level. The scalar $D$ represents the strength of the
Coulomb force acting on a charged grain, $\hat{X}$ is the position
operator, and $x_0 = \sqrt{\hbar / m\omega_0}$ is the harmonic
oscillator length scale for an oscillator with mass $m$ and angular
frequency $\omega_0$.  The third and fourth terms describe the center
of mass movement of the dot in a harmonic oscillator potential so that
$\hat{P}$ is the center of mass momentum operator. The last term is
the tunneling interaction between the lead states and the dot level
and $T_{k,\alpha}(\hat{X})$ is the tunneling coupling strength. We
will consider the case when the tunneling coupling depends
exponentially on the position operator $\hat{X}$, i.e.
$T_{k,R}(\hat{X}) = T_R\exp\{\hat{X}/\Lambda\}$ and $T_{k,L}(\hat{X})
= T_L\exp\{-\hat{X}/\Lambda\}$, where $T_R$ and $T_L$ are constants
and $\Lambda$ is the tunneling length.

To introduce a connection to the quantified vibrational states of the
oscillator we perform a unitary transformation of the
\mbox{Hamiltonian (\ref{eq:hamiltonian})} so that
$UHU^{\dag}=\tilde{H}$, where $U =
e^{\frac{i}{\hbar}\hat{P}d_0c^{\dag}c}$. In this paper we consider the
situation when $\tilde{H}$ has the most symmetric form:
\begin{eqnarray}
    \tilde{H} &=& \sum_{k,\alpha} E_{k,\alpha} a_{k,\alpha}^{\dag}
  a_{k,\alpha} + \tilde{E_0}c^{\dag}c 
  + \hbar\omega_0 \left(b^{\dag}b + \frac{1}{2}\right)
  \nonumber \\
  & & + T_0 \sum_k
  \bigg[ a_{k,R}^{\dag}c e^{x_-b+x_+b^{\dag}} +
  a_{k,L}^{\dag}c e^{-x_+b-x_-b^{\dag}} \bigg]
  + \textrm{h.c.}.
  \label{eq:hamiltonian_2}
\end{eqnarray}
Here $b^{\dag}$ ($b$) is a bosonic creation (annihilation) operator
for the vibronic degree of freedom, and the dimensionless parameters
$x_{\pm} = 1/\sqrt{2}(x_0/\Lambda \pm d_0/x_0)$ (where $d_0=D/(x_0 m
\omega_0^2)$) characterize the strength of the electromechanical
coupling. Furthermore, $T_0$ is the renormalized tunneling coupling
constant, and $\tilde{E_0}=E_0-Dd_0/(2x_0+m\omega_0^2d_0^2/2)$ is the
shifted dot level. For simplicity, but without loss of generality, we
choose $\tilde{E_0}$ equal to the chemical potential of the leads at
zero bias voltage.

First let us study the situation when the mechanical subsystem is
characterized by a low level of excitation. We will consider the case
of small electromechanical coupling. This means that the dimensionless
parameters $x_{\pm}\ll1$ and that only elastic electronic transitions
and transitions accompanied by emission or absorption of a single
vibronic quantum (single-vibronic processes) are important. If the
applied voltage is smaller than $2 \hbar\omega_0 /e$ and the
temperature is equal to zero, the six allowed transitions of this type
are the ones described in \mbox{Fig.~\ref{fig:case12}a}. Here we see
that only elastic tunneling processes and tunneling processes in which
the vibronic degree of freedom absorbs one vibronic quantum are
allowed and as a result the rate equation for the distribution
function of the energy level population $P(n,t)$ has the form:
\begin{eqnarray}
  \Gamma^{-1}\partial_t P(n,t) &=& 
  P(n,t) + (x_+^2+x_-^2)(n+1)P(n+1,t)
  \nonumber \\
  & & - \left( 1 + n(x_+^2 + x_-^2) \right) P(n,t),
  \nonumber
\end{eqnarray}
where $\Gamma=2\pi T_0^2\nu/\hbar$ and $\nu$ is the density of states
in the leads.

It is straightforward to solve these equations and find that the
solution exponentially fast approaches the stable solution $P(0)=1$
and $P(n)=0$ for all $n>0$. As a result, the dimensionless average
extra energy excited in the vibronic subsystem,
\begin{eqnarray}
  E(t) = \sum_{n=0}^\infty nP(n,t), 
\end{eqnarray}
goes to 0.

If the applied bias voltage is increased above the threshold value
\mbox{$V_{c} = 2 \hbar \omega_0 / e$} we instead get the allowed
transitions described in \mbox{Fig. \ref{fig:case12}b}, i.e. two
absorption processes has changed into emission processes where
the energy quantum $\hbar \omega_0$ is transferred to the vibronic
degree of freedom. These transitions lead to the following equation
for $P(n)$: 
\begin{eqnarray}
  \Gamma^{-1}\partial_t P(n,t) &=&
  - \left( 1 + x_-^2 + (n+1)x_+^2 \right) P(n,t)
  \nonumber \\
  & & + x_-^2 (n+1)P(n+1,t) + P(n,t) + x_+^2 n P(n-1,t).
\end{eqnarray}
\begin{figure}[htbp]
  \psfrag{ev}[]{\hspace*{0mm}\raisebox{0mm}{\large{$\mathbf{\frac{eV}{2}}$}}}
  \psfrag{hw}[]{\hspace*{0mm}\raisebox{0mm}{\large{$\mathbf{\hbar \omega_0}$}}}
  \includegraphics[scale=0.35]{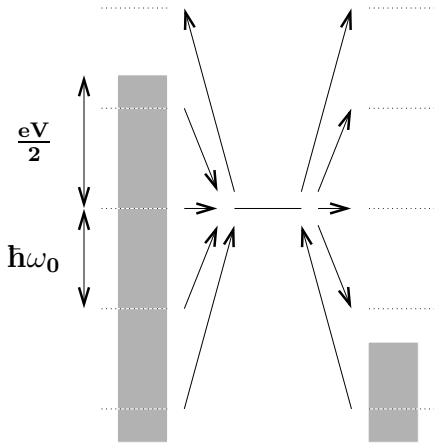}\\
  \vspace{5 mm}
  \caption{ \label{fig:case3} Illustration of the second order case
    where elastic tunneling and inelastic tunneling exchanging two or
    less vibrational quanta are included. The level of the dot is equal
    to the chemical potential and the bias voltage is set so that
    $2\hbar\omega_0 / e < V < 4\hbar\omega_0 / e$.}
\end{figure}

One can find from this equation that the time evolution of the exited
energy is given by the formula:
\begin{eqnarray}
  E(t) = \frac{x_+^2}{x_+^2-x_-^2}
  \left( e^{\Gamma (x_+^2-x_-^2)t}-1 \right),
\end{eqnarray}
i.e. energy is continuously pumped into the mechanical subsystem,
which is strong evidence that the low exited regime is unstable if the
bias voltage exceeds the critical value $V_{c}$. Furthermore, it is
necessary to remark here that for this case we thus have a linear
increase in the energy as a function of time even when $x_+^2$
approaches $x_-^2$.

As the excitation of the vibronic subsystem increases multi-vibronic
processes become important. They give rise to an additional
dissipation which saturates the energy growth induced by the
single-vibronic processes. As a result the system comes to a
stationary regime which is characterized by a significant level of
excitation of the vibronic subsystem. To demonstrate this we will now
expand our analysis by taking into account electronic transitions
accompanied by the emission or absorption of two vibronic quanta
(two-vibronic processes). To describe such transitions one has to take
into account second order terms in $b^{\dag}$ and $b$ in the tunneling
part of the Hamiltonian (\ref{eq:hamiltonian_2}). As illustrated in
\mbox{Fig. \ref{fig:case3}} these terms will generate four processes
in which two vibrational quanta are absorbed by the electron during
the tunneling event. There is also a renormalization of the elastic
channel coming from the inclusion of these terms. Now the equation for
the distribution function of the energy level population has the form:
\begin{eqnarray}
  \Gamma^{-1}\partial_t P(n,t) &=& n P(n-1,t)
  \nonumber \\
  & & - \bigg[ \epsilon n^2
  + (\alpha - \epsilon + 1) n + 1 \bigg] P(n,t)
  \nonumber \\
  & & + \alpha (n+1) P(n+1,t)
  \nonumber \\
  \label{eq:mastereq}
  & & + \epsilon (n+1)(n+2) P(n+2,t)  = 0,
\end{eqnarray}
where we have introduced the constants $\epsilon =
(x_+^4+x_-^4)/(4x_+^2)$ and $\alpha = x_-^2/x_+^2$.

To find the stationary solution of this equation we introduce the
generating function:
\begin{eqnarray}
  {\cal P}(z) = \sum_{n=0}^\infty z^n P(n) \nonumber,
\end{eqnarray}
where $z$ is a complex number inside the unit circle. Rewriting
\mbox{Eq. \ref{eq:mastereq}} we find the equation for ${\cal P}(z)$ 
\begin{eqnarray}
  \epsilon (z+1) \partial_z^2 {\cal P}(z) 
  + (\alpha-z) \partial_z {\cal P}(z)
  - {\cal P}(z) = 0 \nonumber.
\end{eqnarray}
The solution to this equation is
\begin{eqnarray}
  {\cal P}(z) &=& e^{-\int_1^z dz^\prime
    \frac{\alpha-\epsilon-z^\prime}{\epsilon(z^\prime+1)}}
  \nonumber \\
  & & * \Bigg\{
  \int_{z_0}^z dz^\prime
  e^{
    \int_1^{z^\prime} dz^{\prime \prime}
    \frac{\alpha-\epsilon-z^{\prime \prime}}{\epsilon(z^{\prime
        \prime}+1)}
    }
  \frac{C_1}{\epsilon(z^\prime+1)}
  \Bigg\} \nonumber,
\end{eqnarray}
where $C_1$ and $z_0$ are constants. Since the probabilities $P(n)$
are positive and normalized, the sum $\sum_{n=0}^\infty (-1)^n P(n)
= {\cal P}(z=-1)$ converges absolutely. This is true only for $z_0=
-1$. The second constant $C_1$ can be determined from the
normalization condition ${\cal P}(z=1) = 1$ to be $C_1 = \epsilon
2^\gamma/\int_{-1}^1dx e^{(1-x)/\epsilon}(x+1)^{\gamma-1}$, where we
have introduced the constant $\gamma = (\alpha - \epsilon +
1)/\epsilon$. Therefore the final expression for ${\cal P}(z)$ is
\begin{eqnarray}
  {\cal P}(z) = \frac{C_1}{\epsilon}
  \frac{e^{\frac{z-1}{\epsilon}}}{(z+1)^\gamma}
  \int_{-1}^z dz^\prime
  e^{\frac{1-z^\prime}{\epsilon}} (z^\prime+1)^{\gamma-1}.
\end{eqnarray}

We can now calculate the average energy excited in the harmonic
oscillator, which is just $\partial_z {\cal P}(z)$ calculated at
$z=1$,
\begin{eqnarray}
  E = \frac{1}{2\epsilon}(2+C_1-\epsilon\gamma).
\end{eqnarray}

To see how the energy pumped into the harmonic oscillator affects the
charge transport we calculate the current $I$ through the system in
units of $e\Gamma$. For voltages below $V_c$ the current is only
mediated by the elastic channel and is thus $I = 1/2$.
\begin{figure}[htbp]
  \psfrag{ylabel}[]{\hspace*{0mm}
    \raisebox{10mm}{\large{$\mathbf{I}$}}}
  \psfrag{xlabel}[]{\hspace*{4mm}
    \raisebox{-10mm}{\large{$\mathbf{eV/\hbar\omega_0}$}}}
  \includegraphics[scale=0.4]{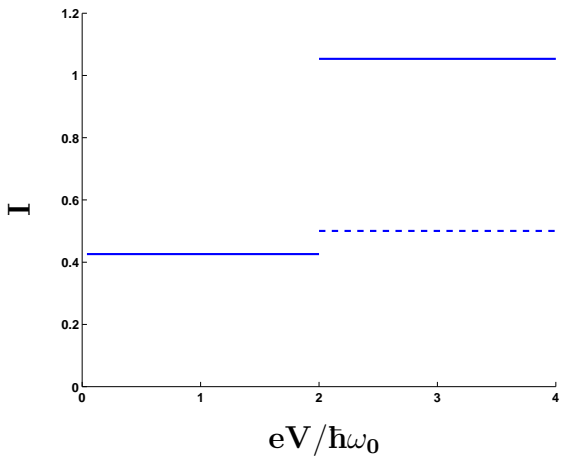}\\
  \vspace{5 mm}
  \caption{\label{fig:current}The current as plotted as a function of
    the bias voltage (solid line). The current makes a jump as the
    non-elastic channel is opened at $V=2\hbar\omega_0/e$. We have
    also plotted (dashed line) the current as a function of voltage
    for the high dissipation limit where the harmonic oscillator goes
    to the ground state between tunneling events.}
\end{figure}

For voltages in the range $2\hbar\omega_0/e < V < 4\hbar\omega_0/e$
the current can be calculated to leading order in $\epsilon$ as 
\begin{eqnarray}
  \label{eq:curr_high}
  I = \left( \frac{1}{4}(x_-^4-x_+^4) + x_+^2x_-^2 \right) 
  {\cal P}^{\prime\prime}(z=1) 
  +  \left(x_+ - x_-\right)^2 {\cal P}^\prime(z=1) + 1.
\end{eqnarray}
In \mbox{Fig. \ref{fig:current}} we have chosen a set of numerical
values and plotted (solid line) the calculated current as a function
of the bias voltage. For comparison we have also plotted (dashed line)
the current as given in the high dissipation limit where the harmonic
oscillator goes to the ground state between tunneling events. It is
clear that the current in the regime characterized by a high level of
excitation is much greater than the one in the regime of low level of
excitation.

In conclusion, we have studied fully quantized mechanical motion of a
single-level quantum dot coupled to two voltage biased electronic
leads. We have shown that above a certain threshold voltage the energy
accumulated in the mechanical subsystem increases dramatically. We
have also shown that second order inelastic tunneling events are
enough to stabilize this pumping of energy. Finally the current
through the system was calculated and it was found that the
development of the mechanical instability is accompanied by a dramatic
increase in the current.

The authors would like to thank Robert Shekhter, Jari Kinaret and
Anatoli Kadigrobov for valuable discussions related to this
manuscript.

\bibliographystyle{plain}

\end{document}